\pdfoutput=1

\documentclass[conference,a4paper]{APSIPA2021}
\usepackage{amsmath}
\usepackage{graphicx}
\usepackage{multirow}
\usepackage{booktabs}
\usepackage{threeparttable}
\usepackage{tablefootnote}

\usepackage[style=numeric, backend=bibtex,style=ieee,]{biblatex}
\bibliography{mybib.bib}

\usepackage{geometry}
\usepackage{fancyhdr}

\fancypagestyle{firststyle}{
  \fancyhf{}
  \fancyhead[C]{2025 Asia Pacific Signal and Information Processing Association Annual Summit and Conference (APSIPA ASC)}
}

\begin{document}

\title{MixedG2P-T5: G2P-free Speech Synthesis for Mixed-script texts using Speech Self-Supervised Learning and Language Model}


\author{Joonyong Park$^*$, Daisuke Saito$^*$, Nobuaki Minematsu$^*$\\

\authorblockA{
\authorrefmark{1}
The University of Tokyo, Japan \\
E-mail: \{jpark, dsk\_saito, mine\}@gavo.t.u-tokyo.ac.jp}
}

\maketitle
\thispagestyle{firststyle}
\pagestyle{fancy}

\begin{abstract}
This study presents a novel approach to voice synthesis that can substitute the traditional grapheme-to-phoneme (G2P) conversion by using a deep learning-based model that generates discrete tokens directly from speech. Utilizing a pre-trained voice SSL model, we train a T5 encoder to produce pseudo-language labels from mixed-script texts (e.g., containing Kanji and Kana). This method eliminates the need for manual phonetic transcription, reducing costs and enhancing scalability, especially for large non-transcribed audio datasets. Our model matches the performance of conventional G2P-based text-to-speech systems and is capable of synthesizing speech that retains natural linguistic and paralinguistic features, such as accents and intonations.\\
\end{abstract}
\vspace{-1mm}

\section{Introduction}
\vspace{-1mm}

Speech synthesis refers to the technology by which machines automatically generate speech audio signals and is commonly known as text-to-speech (TTS). With the advancement of deep learning, speech synthesis models have demonstrated performance that significantly surpasses traditional methods~\cite{borgholt2022brief}. These models typically convert input text into acoustic feature vectors through an encoder, and subsequently generate Mel-spectrograms using techniques such as attention mechanisms or variational inference, which are then transformed into speech by a vocoder~\cite{shen2018natural,kim2021conditional}. The model learns the correspondence between audio samples and their respective “input representations.”

Constructing such deep learning-based speech synthesis systems requires accurately labeled data corresponding to spoken utterances. In conventional approaches, phonemes are typically generated from sample text using grapheme-to-phoneme (G2P) conversion, which are then input into the speech synthesis model. In the case of Japanese, where texts often contain a mix of kanji and kana, phonemes are generated from the mixed-script input, which are subsequently used to synthesize speech. Specifically, some methods rely on rule-based systems to assign required TTS information—such as accent and prosody—based on morphological analysis, while others adopt neural G2P models using CTC or encoder-decoder structures to model the alignment between text and phoneme sequences of differing lengths.

Such transcription tasks are largely conducted manually. While it is possible to incorporate additional information—such as accents and syllable durations—by referring to pronunciation or accent dictionaries, two major challenges remain in building G2P systems. The first is the cost associated with data construction. Generating phonetic elements requires various resources, including pronunciation and accent dictionaries and linguistic rules. Since these supplementary inputs cannot be derived solely from raw text, they must be individually integrated into the system, thereby incurring high annotation costs.

The second challenge lies in the limited support for multilingual text. When dealing with texts that include multiple languages, it becomes difficult for a single G2P model to provide adequate coverage. This necessitates the development of separate models for each language, which further increases costs. Additionally, pronunciation errors can arise when the same character is pronounced differently depending on the language, presenting a significant difficulty in multilingual speech synthesis.

To address these challenges, this study aims to develop a G2P-free, multilingual-capable speech synthesis model by utilizing discrete representations derived from speech self-supervised learning (SSL) models. To achieve this goal, the following key aspects are investigated.

\begin{itemize} \item Performance comparison between input representation using SSL model and conventional G2P representation \item Implementation of G2P for mixed Kanji and Kana situations using discrete representation \end{itemize}

\section{Related Works}
\vspace{-1mm}

\begin{figure}
    \centering
    \includegraphics[width=0.98\linewidth]{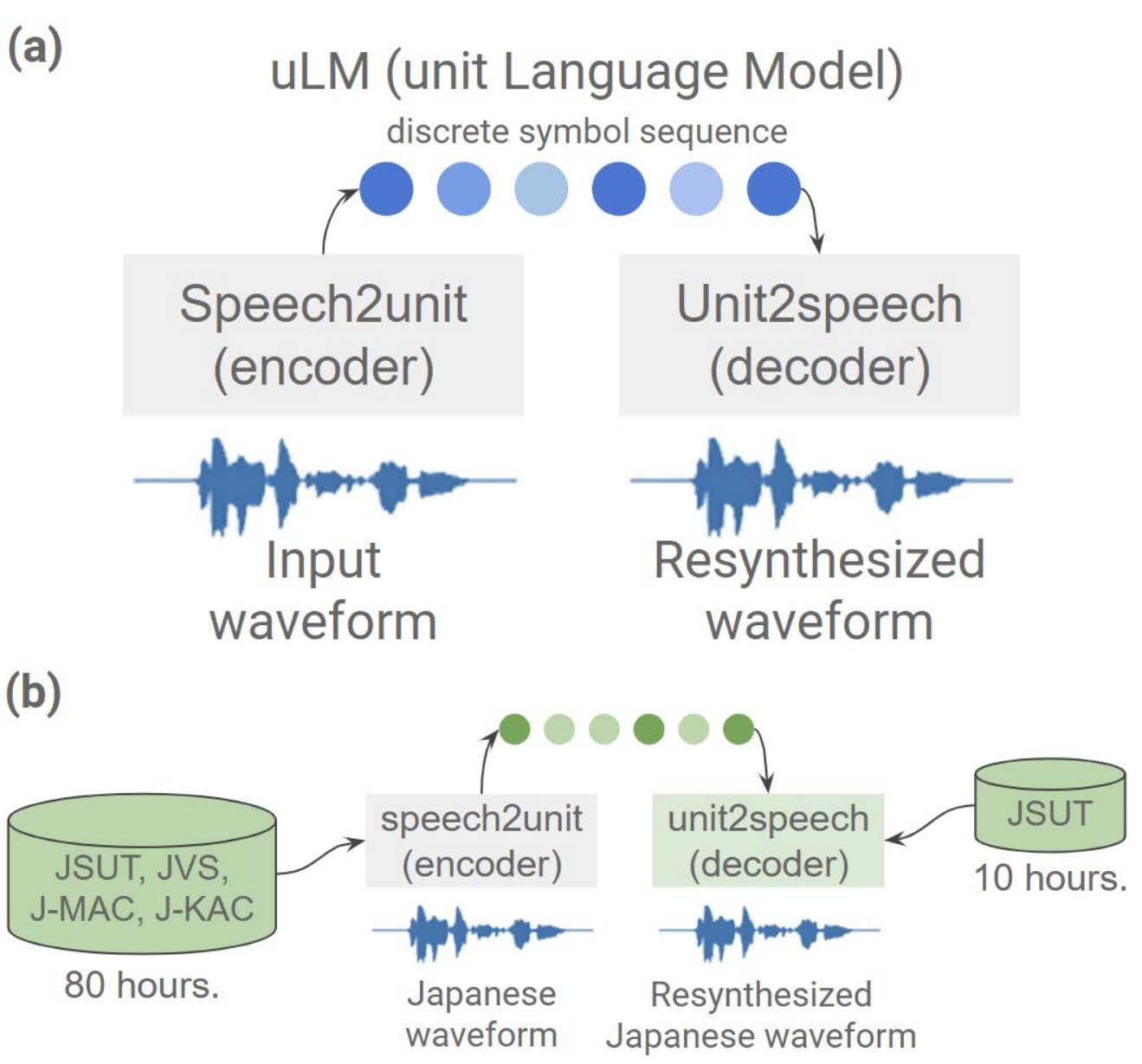}
    \caption{(a) Architecture of GSLM and (b) Application to the Japanese language}
    \vspace{-2mm}
\end{figure}

To address the aforementioned challenges in speech synthesis—particularly the reliance on costly and language-dependent grapheme-to-phoneme (G2P) conversion—recent studies have explored the use of discrete tokens, which are directly extracted from raw audio using speech self-supervised learning (SSL) models trained on large-scale unlabeled speech corpora. Unlike conventional phoneme representations, these discrete tokens encode not only linguistic and semantic content, but also speaker-specific paralinguistic features such as accentuation, intonation, and prosody. This rich and compact representation is especially advantageous because it avoids the information loss typically incurred during the intermediate conversion of speech into text.

A prominent framework that leverages this representation for speech synthesis is the Generative Spoken Language Model (GSLM)~\cite{lakhotia2021generative}. GSLM establishes a three-stage pipeline: (1) an encoder that converts the input speech waveform into a sequence of discrete symbols via quantized SSL embeddings, (2) an optional unsupervised language model (uLM) that captures long-range dependencies across the symbol sequence, and (3) a decoder that reconstructs the speech waveform from the symbolic input. Encoders such as wav2vec 2.0~\cite{baevski2020wav2vec} and HuBERT~\cite{hsu2021hubert} are commonly employed, offering robust and generalizable speech representations. These discrete tokens act as a learned alternative to phonemes or graphemes, forming a flexible intermediate layer between raw audio and synthesis.

As depicted in Figure~1(a), the encoder module transforms the continuous speech signal into a compressed symbolic form using a k-means clustering layer trained on SSL features. The decoder module, typically a neural vocoder or sequence-to-sequence model such as Tacotron 2~\cite{shen2018natural}, learns to reconstruct the speech waveform from these symbolic representations. Interestingly, it has been demonstrated (Figure~1(b)) that this symbolic representation is not strictly language-specific: once the encoder-decoder pipeline is trained on one language, it can be transferred to other languages through fine-tuning, enabling zero-shot or low-resource language synthesis. This cross-lingual transferability opens the door to G2P-free synthesis across many languages, even those lacking well-defined phonemic resources.

\section{Research Approach}

\begin{figure}
    \centering
    \includegraphics[width=0.98\linewidth]{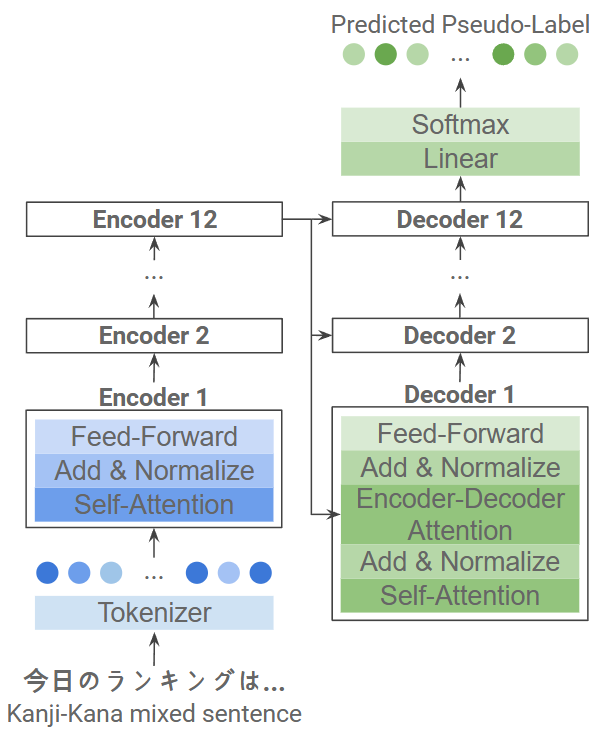}
    \vspace{2mm}
    \caption{Architecture of the Pseudo-Language Label Prediction Model}
    \vspace{-4mm}
\end{figure}

\begin{table*}[tb]
  \centering
  \caption{Datasets used for training, fine-tuning, and applying each model}
  \label{tab:corpus}
  \vspace{2mm}
  \begin{tabular}{ccccc}
    \toprule
    Language & G2P & Language-Specific Pseudo-Language Label Predictor & Spectral Predictor \\\midrule
    Japanese &
    \begin{tabular}{@{}c@{}}
        OpenJTalk\footnotemark[1]\\
        + Mecab\footnotemark[2]\\
        + Marine~\cite{park22b_interspeech}
    \end{tabular} & 
    \begin{tabular}{@{}c@{}}
      Reazonspeech~\cite{reasonspeech} (T5~\cite{Raffel2019ExploringTL} Pre-training)\\ 
      JSUT~\cite{DBLP:journals/corr/abs-1711-00354}, JVS~\cite{DBLP:journals/corr/abs-1908-06248}, JKAC~\cite{jkac}, JMAC~\cite{DBLP:conf/interspeech/TakamichiNTS22}, 
      JSSS~\cite{DBLP:journals/corr/abs-2010-01793} (Fine-tuning for k-means/T5)\\
      tohoku-BERTv3-tokenizer\footnotemark[3] (T5 Tokenizer)
    \end{tabular} &
    JSUT~\cite{DBLP:journals/corr/abs-1711-00354} \\
    \bottomrule
  \end{tabular}

\end{table*}

In this work, we extend this idea by using the encoder of GSLM not only for decoding speech but also for constructing a pseudo-language label predictor, which can infer symbolic representations directly from raw text. The goal is to replace the G2P module with a learned mapping from text, which may contain mixed scripts, such as Kanji and Kana in Japanese, to symbolic tokens that function analogously to phonemes. Once these pseudo-language labels are predicted, they can be fed into the spectral predictor to produce natural-sounding synthesized speech.

The pseudo-language label predictor is built on top of the T5 architecture~\cite{Raffel2019ExploringTL}, a powerful Transformer-based sequence-to-sequence model known for its success in a wide range of natural language processing tasks. T5’s encoder-decoder structure allows it to handle inputs and outputs of varying lengths, making it suitable for converting raw text into sequences of symbolic labels. Moreover, T5 is pretrained on a massive multilingual corpus, which equips it with broad linguistic understanding and generalization capabilities. In our setting, it enables the conversion of Japanese text—including diverse orthographic patterns—to meaningful discrete symbol sequences. The architecture of this label predictor is illustrated in Figure~2. By leveraging this T5-based predictor, we eliminate the need for hand-crafted phoneme dictionaries, morphological analyzers, or accent dictionaries typically required in G2P pipelines.

While prior studies have explored the use of T5 for grapheme-to-phoneme conversion tasks, they remain constrained to alphabetic scripts and conventional G2P paradigms~\cite{rezackova21_interspeech, ao2022speecht5}. Specifically, their work assumes a clear one-to-one correspondence between characters and phonemes, and does not address the combined demonstration of mixed-script languages. Also, multilingual G2P systems have also been proposed, but they still operate within a text-based inference thus does not have additional prosodic features essential for natural-sounding speech synthesis, such as accents and durations, that can be obtained from SSL models~\cite{sokolov19_interspeech, vesik-etal-2020-one, zhu22_interspeech}.

Therefore, this study demonstrate a fully G2P-free speech synthesis pipeline that leverages SSL-derived tokens and a T5-based pseudo-language label predictor to handle the structural and phonological intricacies of mixed-script languages, such as Japanese kanji and kana. 

\section{Experimental Setup}
\vspace{-1mm}

\begin{figure}[tb]
    \centering
    \includegraphics[width=0.98\linewidth]{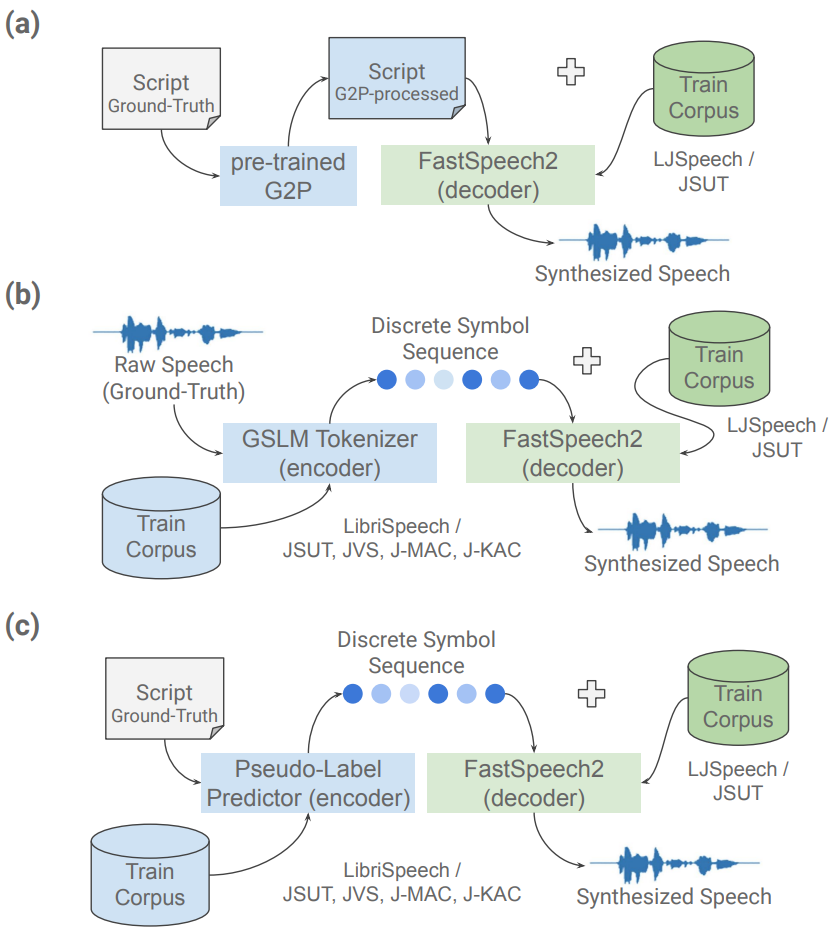}
    \vspace{-2mm}
    \caption{Construction of speech synthesis systems using (a) phonemes obtained via G2P (Baseline), (b) pseudo-language labels obtained from original speech (Oracle), and (c) pseudo-language labels predicted from raw text (Proposed)}
    \label{fig:fig/ronbun_gslm_config2.pdf}
\end{figure}

In order to verify performance of the pipeline, we designed three pipelines based on different encoding architectures to process Japanese sentences. The overall structure is illustrated in Figure 3.\footnotetext[1]{https://open-jtalk.sp.nitech.ac.jp/}\footnotetext[2]{ \url{http://taku910.github.io/mecab/}}\footnotetext[3]{\url{https://huggingface.co/tohoku-nlp/bert-base-japanese-v3}}

The first structure serves as the conventional baseline, using G2P-derived representations obtained from reference scripts. This G2P system, trained using traditional methods such as HMMs, provides phoneme sequences along with information on duration and accent, which are used as inputs for speech synthesis.

The second structure serves as the oracle and utilizes pseudo-language labels derived directly from raw speech using the encoder of the GSLM system trained in a self-supervised manner. Since GSLM does not support synthesis from text scripts, this model generates labels from the original speech aligned with the reference scripts. These labels are then used to evaluate the performance of the pseudo-language label predictor, which is trained on the corresponding discrete representations.

The third structure is our proposed method, which employs a pseudo-language label predictor to infer labels from raw text. Discrete labels are first extracted from raw speech using the GSLM encoder, and a dataset of text–label pairs is constructed using the corresponding reference scripts. This dataset is then used to train the predictor model, enabling it to output predicted pseudo-language labels given new text input.
The pseudo-language label predictor is trained on Japanese utterances using a T5 model with 12 encoder and decoder layers, pre-trained and fine-tuned on TTS data (see Table 1).

oracle and the proposed method both employ the GSLM architecture enhanced with speaker embedding capabilities using ContentVec~\cite{qian22b}, an SSL model known for effectively capturing speaker characteristics. The SSL and k-means models were fine-tuned on a Japanese speech corpus, resulting in a model capable of outputting 500 discrete tokens. In both cases, a preprocessing step was applied to remove repeated symbols from the pseudo-language label sequences.

The spectral predictor used to generate synthesized speech from the common input representations is FastSpeech 2~\cite{Ren2020FastSpeech2F}. FastSpeech 2 has the advantage of being able to additionally control speech characteristics through duration, pitch, and accent information, in addition to textual inputs. The model is trained to predict Mel-spectrograms from input text, which are then converted to speech using a vocoder under the same conditions.

\begin{table*}[t]
    \centering
    \caption{Evaluation metrics obtained from the discrete symbols generated by the language label predictor and synthesized speech generated by the spectral predictor}
    \vspace{2mm}
    {
        \begin{tabular}{cc|c|c|c|c}
        \toprule
            & & GT & Baseline & Oracle & Proposed  \\ 
            \midrule
            Discrete Symbols & {UER(\%)$\downarrow$} &  &  & - & 7.47 \\
            \midrule
            & {CER(\%)$\downarrow$} & 16.21  & 18.24 & 20.63 & 21.28 \\
            Synthesized Speech & {UTMOS$\uparrow$} & 2.79 & 2.59 & 2.49  & 2.54  \\
            (Spectral Predictor) & {WARP-Q$\uparrow$} & - & 2.47 & 2.64  & 2.63   \\
            & {SDR(dB)$\uparrow$} & - & -22.79& -23.12 & -23.28 \\
            \bottomrule
        \end{tabular}
    }
\end{table*}

In the conventional approach, the model is trained on paired data consisting of speech and G2P-derived text. Here, phoneme sequences, duration, and accent information are learned using pre-constructed TextGrid files based on G2P processing.

In the oracle and proposed methods, training is conducted using pairs of speech and pseudo-language labels. Since the output pseudo-language labels may include repeated symbols, these repetitions can be used as cues to predict duration.

\section{Experimental Results}

We first evaluate the pseudo-language label predictor by assessing the linguistic information captured in the predicted discrete symbols. Then, we evaluate the synthesized speech generated via the spectral predictor. Table~2 summarizes the evaluation metrics across the three model architectures, using 100 utterances selected from the JVS corpus that were not used during training.

\subsection{Evaluation of Linguistic Intelligibility} \label{ssec:ryo}

To assess the impact of each input representation on linguistic intelligibility, we analyzed error rates with respect to the target language. For the discrete symbols, we compared the pseudo-language labels predicted by the label predictor with those obtained from the original audio via the SSL model, and calculated the Unit Error Rate (UER). For the synthesized speech, we used the Whisper-base~\cite{radford2022whisper} ASR model and compared its output with the reference transcription, calculating the Character Error Rate (CER) in Japanese.

Table~2 shows the UER and CER scores. While UER did not show large differences compared to prior experiments, prediction errors were still observed in the label predictor. For the synthesized speech, although the CER increased compared to the G2P-based baseline, the increase was smaller when compared to the oracle model. This suggests that synthesis errors are more heavily influenced by the spectral predictor than by the label predictor.

\subsection{Evaluation of Naturalness}

We employed the pretrained UTMOS model~\cite{DBLP:conf/interspeech/SaekiXNKTS22} for this evaluation. UTMOS is a model trained to predict Mean Opinion Scores (MOS) for multilingual synthetic speech in an automated fashion. Since it operates without comparison to reference audio, it can be used to assess absolute linguistic and paralinguistic qualities, though it cannot capture comparative variations between ground truth and synthesized audio.

Table~2 also includes UTMOS scores. While the baseline exhibited the highest naturalness, the proposed method achieved a higher UTMOS score than the oracle model. This further supports the finding that differences in naturalness between the baseline and proposed method are greater than those between oracle and proposed methods, indicating that the spectral predictor has a more substantial influence on naturalness than the label predictor.

\subsection{Evaluation of Acoustic Quality}

Next, we evaluated the overall acoustic quality of the synthesized speech.

Given that input representations are compressed speech features, they can be viewed as analogous to neural audio codecs. Thus, we treated the speech synthesis system as a single transmission pipeline and evaluated the output speech as a degraded version of a reference signal.

For this purpose, we employed WARP-Q~\cite{Wissam_IET_Signal_Process2022}, which is robust to common codec-related distortions and compensates internally for temporal misalignments between the ground truth and resynthesized signals—issues often encountered with metrics like PESQ~\cite{pesq}, which can be sensitive to speaker identity and speech type.

We also assessed audio quality in terms of noise contamination using the Signal Distortion Rate (SDR), a metric typically used in source separation. SDR is suitable here as it quantifies the degree of noise in the synthesized audio relative to the original input.

As shown in Table~2, WARP-Q scores for the proposed model were higher than those of the baseline and comparable to the oracle model. Similarly, SDR results were consistent across the baseline, oracle, and proposed models. These results suggest that the overall pipeline involving the label predictor does not introduce significant acoustic degradation.


\section{Conclusion}
\vspace{-1mm}

This study presented a novel approach to speech synthesis that replaces conventional grapheme-to-phoneme (G2P) conversion by employing a model that directly generates discrete tokens from speech. Furthermore, we trained a pseudo-language label predictor and a spectral predictor using a speech SSL model on mixed Kanji-Kana Japanese texts and analyzed the outputs under various conditions. Experimental results demonstrated that it is possible to predict pseudo-language labels from text with comparable performance to traditional G2P-based models.

Several areas for improvement have been identified. For the language label predictor, we utilized a Japanese tokenizer pretrained on Japanese text. However, to support multilingual settings, it is necessary to explore alternative tokenizers such as Byte-Pair Encoding (BPE) tokenizers that do not rely on language-specific training. For the spectral predictor, additional acoustic features—such as accentual information—should be integrated alongside the four currently used features.

From an evaluation perspective, it is also necessary to individually assess the contributions of each input factor (e.g., duration, pitch) fed into FastSpeech 2. Moreover, a broader and more diverse test dataset would allow for a more comprehensive and generalized analysis.

As future directions, we plan to conduct experiments on multilingual datasets to investigate the language dependency of the language label predictor. In order to minimize such dependency, we intend to explore strategies that avoid language-specific preprocessing. One possible solution is to convert raw text into language-aware discrete tokens using tokenizers such as mT5~\cite{mt5cite} or ByT5~\cite{byt5cite}, which tokenize text based on Unicode character strings. These models are capable of performing tokenization while preserving minimal language-specific biases, thus enabling scalable and robust evaluation across a wider range of multilingual scenarios.

\printbibliography

\end{document}